\documentclass[a4, 11pt]{article}
\usepackage{amsmath} 
\usepackage{amssymb} 

\usepackage{geometry} 
\usepackage{graphicx} 
\usepackage{epsfig}

\usepackage{booktabs} 
\usepackage{array} 
\usepackage{paralist} 
\usepackage{verbatim} 
\usepackage{subfig} 
\usepackage[dutch]{babel}
\usepackage{fancyhdr} 
\usepackage{sectsty}
\allsectionsfont{\sffamily\mdseries\upshape} 

\usepackage[titles,subfigure]{tocloft} 

\title{MPI-based Evaluation of Coordinator Election Algorithms}
\author{Filip De Turck, Ghent University - imec, Belgium}   
\date{}
\begin{document}
\maketitle
\renewcommand{\abstractname}{Abstract}
\renewcommand{\refname}{References}
\begin{abstract} 
In this paper, we detail how two types of distributed coordinator election algorithms can be compared in terms of performance based on an evaluation on the High Performance Computing (HPC) infrastructure. An experimental approach based on an MPI (Message Passing Interface) implementation is presented, with the goal to characterize the relevant evaluation metrics based on statistical processing of the results. The presented approach can be used to learn master students of a course on distributed software the basics of algorithms for coordinator election, and how to conduct an experimental performance evaluation study. Finally, use cases where distributed coordinator election algorithms are useful are presented.
\end{abstract}

\section{Problem statement}
The coordinator election algorithms are stated as follows. We consider a set of N processes ${p_1, . . . , p_N}$ that (i)~have no shared variables, and (ii)~can communicate by message passing. From this set, we want to select one process to play a special role (e.g. coordinator), such that (i)~every process $p_i$ should have the same coordinator, (ii)~if the elected process fails a new election round is started. In order to be able to make this choice, we assume that each process has a unique ID, and the problem is to elect the process with the largest ID. This ID could e.g. be related to the inverse of the load on the process, in order to elect a relatively unloaded process. This problem statement implies that IDs are unique and that IDs are totally ordered.

\section{Evaluation metrics}
There are two important metrics to characterize the performance of algorithms for coordinator election:
\begin{itemize}
\item Bandwidth: the number of messages used to elect a new coordinator. 
\item Turnaround time: time needed for the election (measured from the moment the election process is started until all processes have agreed on a common coordinator). This turnaround time is typically expressed as a function of $\delta$, the one-way communication delay between process nodes).
\end{itemize}

\section{Two considered algorithms}
\subsection{Chang-Roberts Algorithm}
The Chang-Roberts election algorithm assumes that all processes are organized in a logical ring, where each process can communicate with its successor in the ring. Each process has a unique ID, and the purpose of the algorithm is to elect the process with the largest ID as coordinator. The algorithm assumes reliable communication channels between live processes (no crashes while the election is ongoing). No assumptions are made on channel or processing delays, so the algorithm supports asynchronous systems.

\subsection{Franklin Algorithm}
This election algorithm is a variant of the Chang-Roberts algorithm. It is based on bidirectional communication in a ring, whereas the Chang-Roberts algorithm is based on unidirectional communication in a ring. Nodes are either active or passive. Initially, all nodes are active. \\
The algorithm works in multiple rounds. In each round, each active process takes the following actions: (i)~ it sends the token with process ID to both neighbours, (ii)~ it examines the two tokens received from its neighbours: if one or two tokens contain ID $>$ process ID, the process turns passive, i.e. only forwards tokens to neighbours (i.e. acts as a router), and (iii)~when there is only one active process left (i.e. process receives its own tokens), this process becomes the coordinator.\\
In a ring with N active processes, at least $N/2$ turn passive after one round. After at most ${log}_2 N$ rounds, there is only one active process left. Then one additional round is needed to become leader. In total, $1 + {log}_2 N$ rounds are needed to elect an unique coordinator.

\section{Approach}
Consider the following two algorithms for distributed coordinator election:
\begin{enumerate}

\item implement the Chang-Roberts algorithm based on MPI for execution on the HPC infrastructure. Make sure to keep track of the amount of messages that were sent by the algorithm. Run the solution 10 times, each time using 16 processes. Give the minimum, maximum and average (averaged over 10 runs) amount of messages that your solution needs to elect a coordinator. In addition, also give the minimum, maximum and average (averaged over 10 runs) execution times.

\item implement the Franklin algorithm based on MPI for execution on the HPC infrastructure. Make sure to keep track of the amount of rounds that were needed to elect a coordinator by the algorithm. Run the solution 10 times, each time using 16 processes. Give the minimum, maximum and average (averaged over 10 runs) number of rounds that your solution needs to elect a coordinator. In addition, also give the minimum, maximum and average (averaged over 10 runs) execution times.

\end{enumerate}

Assume $N$ equals 16, implement the two algorithms above in the C++ programming language and by making use of the MPI (Message Passing Interface) library. Determine the used bandwidth and obtained turnaround times for the two algorithms.

\section{HPC infrastructure}
All code is compiled and executed using the UGent HPC (High Performance Computing) infrastructure. Once connected to the HPC infrastructure, students are logged on to one of the interface nodes. These nodes can be used to compile software and submit jobs to the different clusters, but running software on these interface nodes is generally considered bad practice. For this experimental evaluation, however, it is fine to run lightweight examples on the interface node. Hence, you may simply execute the commands in the assignments on the interface nodes. The project should be compiled and run on a interface nodes. In order to compile all source files, please use the provided compile.sh script (that relies on CMake internally to generate a Makefile). You can also change the number of parallel processes MPI should start, which is specified by \texttt{-np}.

\section{Discussion}
The comparison study of the two algorithms for coordinator election contributes to acquiring knowledge and getting acquainted with:

\begin{enumerate}

\item the internal details of both algorithms (both ring-based, one algorithm is an extension of the other),

\item the typical communication delays and processing delays obtained on the High Performance Cluster,

\item the experimental evaluation setup on the High Performance Cluster and statistical processing of the obtained numerical results, and

\item typical numerical values for the two important evaluation metrics: bandwidth and turnaround time.

\end{enumerate}

Coordination election algorithms are very useful and required in several application domains. Important use cases where coordinator election algorithms can be applied to optimize the system behaviour are: 
\begin{itemize}

\item hierarchical network management systems~\cite{hiermgmtsystem, hierMoens}, where the coordinator can collect additional data and initiate the decision making process,

\item softwarized network management~\cite{vnfp, tnsmmoens}, where the coordinators can contribute to the dynamic resource allocation decisions,

\item adaptive video delivery~\cite{scalablevideolaga, 6dofadaptive, reviewMariaQoE}, where the coordinators estimate the most appropriate video quality levels and inform the other involved nodes,

\item efficient virtual desktop cloud computing~\cite{SupercomputingDeboosere}, where coordinators in a hierarchical setting can optimize the resource allocation process,

\item smart city applications~\cite{CNSM2017Santos, NOMS2018Santos} and mobile augmented reality applications~\cite{JSSVerbelen}, where the coordinator can collect and aggregate additional data for informing the other nodes,

\item replica placement in ring based content delivery networks~\cite{ComComWauters}, where coordinator election algorithms are important to generate a hierarchical and scalable operation,

\item optimization the delivery of adaptive video streaming services by in network optimizations~\cite{ToMBouten, ComNetBouten}.

\end{itemize}

\section{GitHub repository}
\texttt{https://github.ugent.be/fdeturck/PDS/tree/main/HWA5}


\begin{thebibliography}{9}

\bibitem{hiermgmtsystem} J. Famaey, S. Latr\'{e}, J. Strassner, F. De Turck, A hierarchical approach to autonomic network management, 2010 IEEE/IFIP Network Operations and Management Symposium Workshops, IEEE NOMS 2010, Osaka, Japan, pp 225-232.

\bibitem{hierMoens} H Moens, B Hanssens, B Dhoedt, F De Turck, Hierarchical network-aware placement of service oriented applications in clouds, IEEE Network Operations and Management Symposium, IEEE NOMS 2014, Cracow, Poland, pp. 1-8.

\bibitem{vnfp} H. Moens, F. De Turck, VNF-P: A model for efficient placement of virtualized network functions, IEEE International Conference on Network and Service Management (CNSM), 2014, Rio De Janeiro, Brazil, pp. 418-423.

\bibitem{tnsmmoens} H. Moens, F. De Turck, Customizable function chains: Managing service chain variability in hybrid NFV networks, IEEE Transactions on Network and Service Management 13 (4), 2016, pp. 711-724.

\bibitem{scalablevideolaga} S. Laga, T. Van Cleemput, F. Van Raemdonck, F. Vanhoutte, N. Bouten, M. Claeys, F. De Turck, Optimizing scalable video delivery through OpenFlow layer-based routing, 2014 IEEE Network Operations and Management Symposium (NOMS), IEEE NOMS 2014, pp. 1-4.

\bibitem{6dofadaptive} J. van der Hooft, T. Wauters, F. De Turck, C. Timmerer, H. Hellwagner, Towards 6dof http adaptive streaming through point cloud compression, Proceedings of the 27th ACM International Conference on Multimedia, ACM Multimedia, 2019, pp. 2405-2413.

\bibitem{reviewMariaQoE} M. Torres Vega, C. Perra, F. De Turck, A. Liotta, A review of predictive quality of experience management in video streaming services, IEEE Transactions on Broadcasting, Volume 64, Issue 2, IEEE, pp. 432-445.

\bibitem{SupercomputingDeboosere} L. Deboosere, B. Vankeirsbilck, P. Simoens, F. De Turck, B. Dhoedt, P Demeester, Efficient resource management for virtual desktop cloud computing, Springer Journal of Supercomputing 62 (2), 2012, pp. 741-767.

\bibitem{CNSM2017Santos} J. Santos, T. Wauters, B. Volckaert, F. De Turck, Resource provisioning for IoT application services in Smart Cities, IEEE International Conference on Network and Service Management (CNSM), 2017, Tokyo, Japan, pp. 1-9.

\bibitem{NOMS2018Santos} J. Santos, P. Leroux, T. Wauters, B. Volckaert, F. De Turck, Anomaly detection for smart city applications over 5g low power wide area networks, NOMS 2018-2018 IEEE/IFIP Network Operations and Management Symposium, IEEE NOMS 2018, pp. 1-9.

\bibitem{JSSVerbelen} T. Verbelen, T. Stevens, P. Simoens, F. De Turck, B. Dhoedt, Dynamic deployment and quality adaptation for mobile augmented reality applications, Journal of Systems and Software, Volume 84, Issue 11, 2011, Elsevier, pp. 1871-1882.

\bibitem{ComComWauters} T. Wauters, J. Coppens, F. De Turck, B. Dhoedt, P. Demeester, Replica placement in ring based content delivery networks, Elsevier Journal on Computer Communications, Volume 29, Issue 16, 2006, pp. 3313-3326.

\bibitem{ToMBouten} N. Bouten, S. Latr\'{e}, J. Famaey, W. Van Leekwijck, F. De Turck, In-network quality optimization for adaptive video streaming services, IEEE Transactions on Multimedia, 2014, Volume 16, Issue 8, pp. 2281-2293.

\bibitem{ComNetBouten} N. Bouten, R. de O Schmidt, J. Famaey, S. Latr\'{e}, A. Pras, F. De Turck, QoE-driven in-network optimization for Adaptive Video Streaming based on packet sampling measurements, Elsevier Computer Networks, Volume 81, 2015, pp. 96-115.


\end{thebibliography}
\end{document}